\documentclass[11pt]{article}
\usepackage{moriond,epsfig}
\usepackage{amssymb}
\usepackage{amsbsy}
\usepackage{amsmath}
\usepackage{amsfonts}
\usepackage{verbatim}

\def\be{\begin{equation}}
\def\ee{\end{equation}}
\def\bea{\begin{eqnarray}}
\def\eea{\end{eqnarray}}

\begin{document}
\vspace*{4cm}
\title{Has saturation physics been observed in deuteron-gold collisions at RHIC?}

\author{ Yacine Mehtar-Tani}

\address{\it Laboratoire de Physique Th\'eorique, Universit\'e de
Paris XI, \\B\^atiment 210,
91405 Orsay Cedex, France}

\maketitle\abstracts{
We have addressed the question of whether saturation (CGC) has been observed in deuteron-gold collisions at RHIC. We have made a detailed analysis of the Cronin peak characteristic of the nuclear modification factor measured for d-Au
  collisions at mid-rapidity. The Cronin peak which is obtained
 around $p_t\simeq 3$ GeV may be reproduced at the proper
  height only by boosting the saturation momentum by a huge non-perturbative additional 
  component. \\
At forward rapidity, we get a quantitative agreement with data, reproducing hadron production spectra and the $R_{CP}$ ratio using a recently developed description of the small-x physics.\\}
\section{Introduction}

During the last two decades, a very rich activity of both theorists and experimentalists has led to a better understanding of QCD at very high energy and high density. Especially with the advent of RHIC, which provided a very good opportunity to test new ideas. New ideas such as the theory of the Color Glass Condensate (CGC) (see ref. \cite{JKrev} and references listed therein) which describes the physics of saturation in the initial state of heavy ion collisions.   
\section{Hadron production in d-A collisions}

We focus our study on the 
  $R_{CP}$ (Central/Peripheral collisions) 
ratio defined as
\begin{equation}\label{RpA}
R_{CP}=\frac{N^{P}_{coll}\frac{dN^{dA\rightarrow hX}}{d\eta
d^{2}{\bf k} }\vert_{C}}{N^{C}_{coll}\frac{dN^{dA\rightarrow
hX}}{d\eta d^{2}{\bf k}}\vert_{P}}.
\end{equation}
${\bf k}$ and $\eta$ are respectively the transverse momentum and
the pseudo-rapidity of the observed hadron. $N_{coll}$ is the
number of collisions in dA, it is roughly twice the number of
collisions in pA (proton-Gold). The centrality dependence of
$R_{CP}$ is related to the dependence of $N^{dA\rightarrow
hX}=d\sigma^{dA\rightarrow hX}/d^{2}{\bf b}$ and $N_{coll}({\bf
b})$ on the impact parameter of the collision. In this paper, we
address the predictions of the (CGC) for this
ratio. We always assume that cross-sections depend on the impact
parameter only through the number of participants which is
proportional to the saturation scale $Q_ {sA}^2({\bf b})\simeq Q_ {sA}^2(0)N _{part.Au}({\bf b})/N
_{part.Au}(0)$, where $N_{part.Au}$ is the number of participants in the gold
nucleus in d-Au collisions \cite{KLN}. 
 Also, we use Table 2 in ref. \cite{BRAHMS1} which gives the number
of participants $N_{part}$ and the number of collisions $N_{coll}$
for several centralities.
\subsection{mid-rapidity}\label{subsec:prod}
At mid-rapidity, $x_d=x_A=k_\perp/\sqrt{s}\simeq 10^{-2}$; for such small values -but not small enough to include quantum evolution- of the momentum fraction carried by partons in the deuteron and in the nucleus, gluons dominate the dynamics, therefore we focus on gluon production in the semi-classical picture given by the Mueller-Kovchegov formula \cite{KovMul}:
\begin{equation}\label{cross}
\frac{d\sigma^{dA\rightarrow gX}}{d\eta d^{2}{\bf k} d^{2}{\bf
b}}=\frac{C_{F}\alpha_{s}}{\pi^{2}}\frac{2}{{\bf
k}^{2}}\int_{0}^{~1/\Lambda}du\ln\frac{1}{u\Lambda}\partial_{u}[u\partial_{u}N_{G}(u,{\bf
b })]J_{0}(\vert {\bf k}\vert u),
\end{equation}
where $u=\vert {\bf z}\vert$ and $\Lambda\sim\Lambda_{QCD}$ is an infrared cut-off.\\
The CGC approach yields a
Glauber-Mueller form for the gluon dipole forward scattering amplitude
\begin{equation}\label{NG}
N_{G}({\bf z},{\bf b})=1-{\bf exp}({-\frac{1}{8}{\bf
z}^2Q^{2}_{s}({\bf b})\ln\frac{1}{{\bf z}^2\Lambda^2}}).
\end{equation}  
In fig. (1) (data from Ref.\cite{BRAHMS1}) we see that the perturbative estimate for the saturation scale $Q_{sC}^2=2$ GeV$^2$, is not sufficient to reproduce the experimental Cronin peak. Thus, we need to enhance $Q_{s}^2$ by a non-perturbative component\cite{AG,KKT2} up to a value of $Q_{s}^2=9$ GeV$^2$ in order to get a meaningful
comparison with data (we have not included error bars in the figures).
\begin{figure}[h]
\centering
\includegraphics[width=5cm]{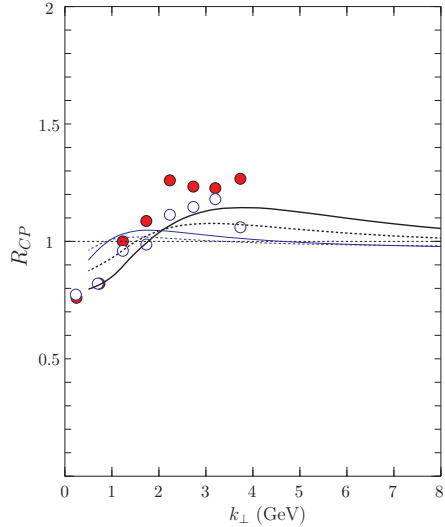}
\caption{ 
$R_{CP}$ for $Q^2_{s.C}=9$ GeV$^2$ (thick lines) and
$Q^2_{s.C}=2$ GeV$^2$ (thin lines). Full lines correspond to
central over peripheral collisions (full experimental dots), dashed lines correspond to semi-central over peripheral collisions
(empty experimental dots).}
\end{figure}
\subsection{Forward rapidity}
At forward rapidity the treatment is different, indeed, $x_d=k_\perp e^\eta/\sqrt{s} \simeq 10^{-1}-1$, thus we treat the parton emerging from the deuteron in the framework of QCD factorization. In the nucleus, $x_A=k_\perp e^ {-\eta}/\sqrt{s}\simeq 10^{-4}-10^{-3}$: gluons dominate and in this very small-x region we should apply $k_t$-factorization. The hadron production cross-section is written as follows: 
\begin{equation}
\frac{d\sigma^{dA\rightarrow hX}}{d\eta d^{2}{\bf k} d^{2}{\bf
b}}=\frac{\alpha_{s}(2\pi)}{ C_{F}}\sum_{i=g,u,d}\int_{z_0}^1
dz\frac{\varphi_{A}({\bf k}/z,Y+\eta+\ln z,b)}{{\bf k}^{2}}[
f_{i}(x_{d}/z,{\bf k}^{2}/z^{2})D_{h/i}(z,{\bf k}^2)],
\end{equation}
where $ f_{u,d}(x,{\bf k}^{2})=(C_{F}/N_{c})xq_{u,d}(x,{\bf
k}^{2})$ and $f_{g}(x,{\bf k}^{2})=xg(x,{\bf k}^{2})$ are the
parton distributions inside the proton; $D_{h/i}(z,{\bf k})$ are
Fragmentation Functions of the parton $i$ into hadron $h$, and
$z_0=(k_{\bot}/\sqrt{s})e^{\eta}$, and $\varphi_A$ is the unintegrated gluon distribution in the nucleus, it is related to the Fourrier transform of the forward dipole scattering amplitude: $\varphi_{A}(L,y)\propto \frac{d^2}{dL^2}\tilde{N}(L,y)$. At large $y$, the BK equation\cite{BK} (derived in the framework of the CGC) provides the following expression \cite{munp2,MuT,IIMc}:

\begin{equation}\label{BKsol}
\tilde{N}(L,Y)\propto L\exp[-\gamma_{s}L-\beta (y) L^{2}].
\end{equation}
It has a remarkable geometric scaling behavior in the variable $L=\ln ({\bf
k}^{2}/Q_{s}^{2}({\bf b},y))+L_0$ when $y$ goes to infinity, $L_0$ is a constant fixed as in \cite{MuT}. 
$\gamma_s\simeq 0.628$ is the anomalous
dimension of the BFKL dynamics in the geometric scaling region
\cite{MuT,IIMc}, and $\beta(y)\propto1/y$. We
used the fit to the HERA data performed in ref.\cite{IIM} in order to fix some free parameters. 
In fig. (2) (data from Ref.\cite{BRAHMS1}), we show our results for $R_{CP}$ for different rapidities. The agreement of the CGC-inspired BK- description at forward rapidity is quite good, even at $\eta=1$, where our approach is however no longer valid.
\begin{figure}[hbtp]
\centering
 \begin{tabular}{c c c }

\qquad\includegraphics[width=4cm]{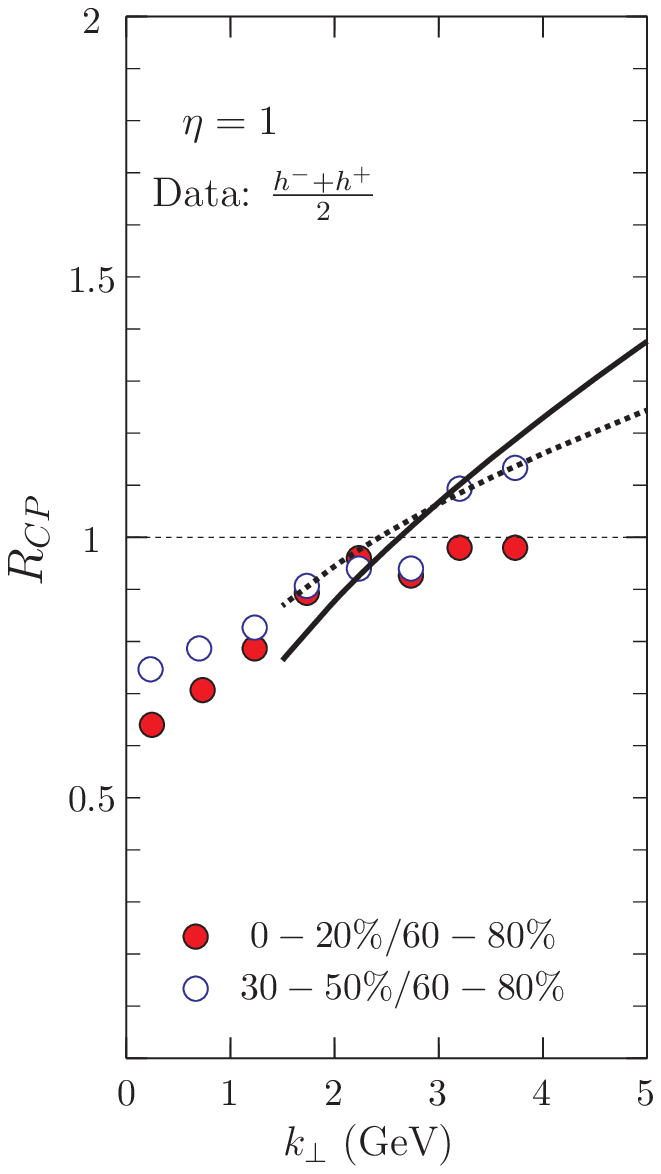} &\qquad   \includegraphics[width=4cm]{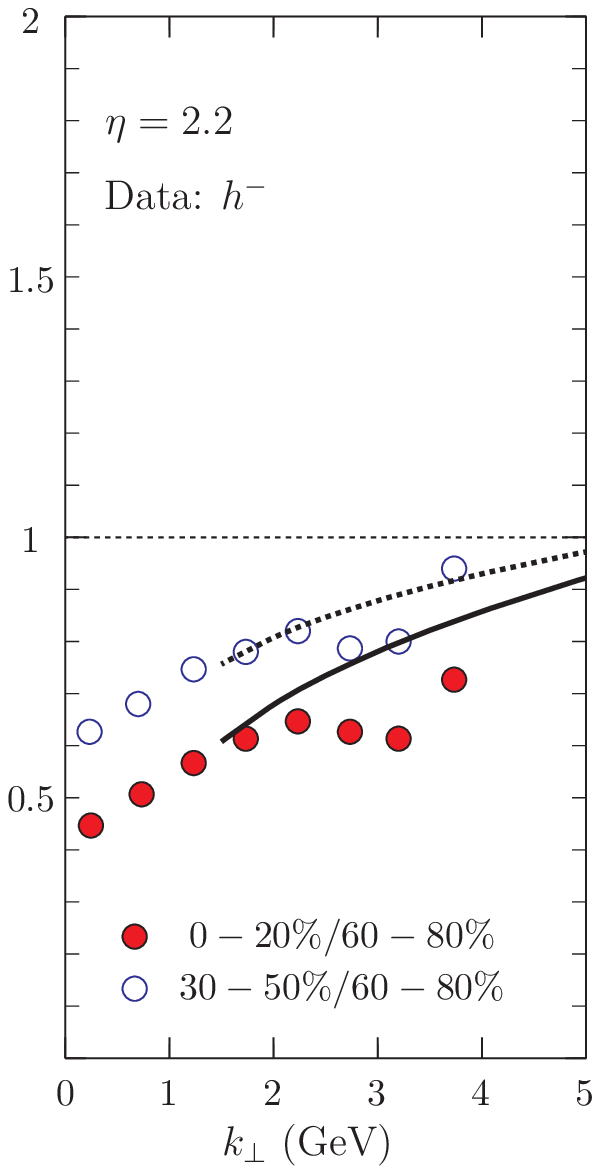} &\qquad   \includegraphics[width=4cm]{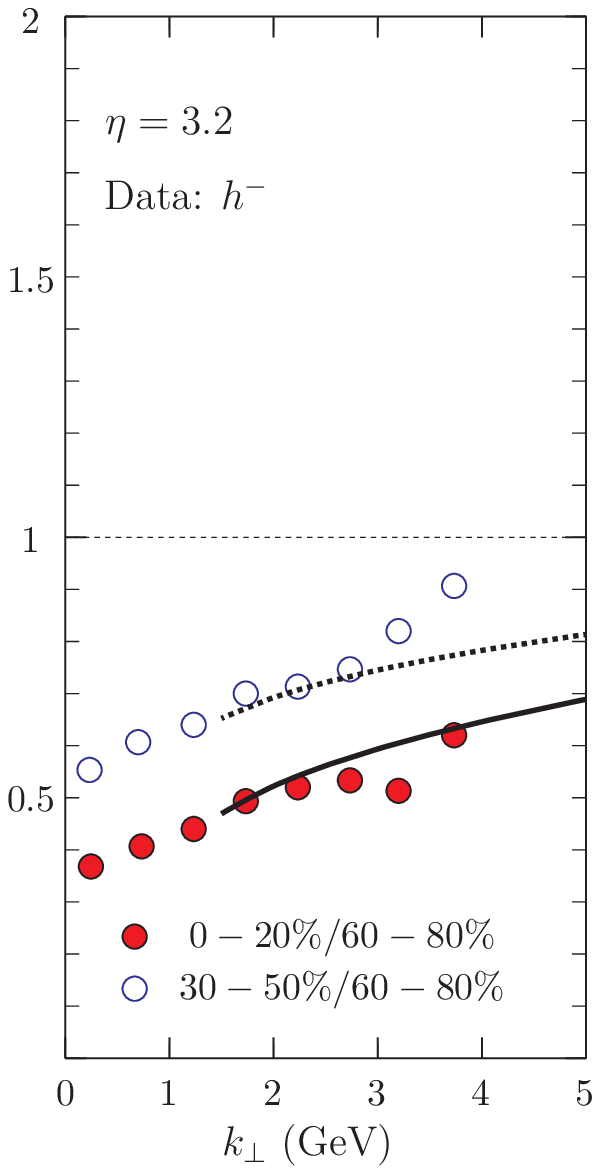}\\
      \qquad (a)& \qquad(b)& \qquad(c)
\end{tabular} 
\caption{
$R_{CP}$ at different rapidities $\eta=1,
2.2$ and $3.2$. Full lines correspond to
central over peripheral collisions (full experimental dots), dashed lines correspond to semi-central over peripheral collisions
(empty experimental dots).} 
\end{figure}

 \section{Summary}
 At mid-rapidity, the CGC is not sufficient to yield a quantitative description of the Cronin peak. Whereas, at forward rapidity, we obtain a good quantitative agreement with data -not only the $R_{CP}$ is reproduced, but also hadron spectra \cite{AY}.  
It should be noticed that the main features of $R_{CP}$ may be in fact understood within the approximate form (when
$k_{\bot}\gtrsim Q_s$)
\begin{equation}
R_{CP}\simeq
\left(\frac{N^C_{part}}{N^P_{part}}\right)^{\gamma_{eff}-1}.
\end{equation}
 At forward rapidity
$\gamma_{eff}\simeq\gamma_s+\beta(\eta)\ln(k_{\bot}^2/Q_s^2)$ is a
decreasing function of $\eta$ and an increasing function of
$k_{\bot}$. This allows us to understand the qualitative behavior
shown by data and in particular the inversion of the centrality
dependence compared to mid-rapidity.
At very large $\eta$ the anomalous dimension stabilizes at
$\gamma_{eff}=\gamma_s$, which could be tested at the LHC.

\end{document}